\documentclass[preprint,12pt,authoryear]{elsarticle}
\usepackage{amsmath,amssymb}
\usepackage{hyperref}
\usepackage{booktabs}

\usepackage{graphicx,caption,subcaption}
\usepackage{natbib}
\usepackage{multirow}
\usepackage{lineno}
\usepackage{aas_macros}
\modulolinenumbers[5]

\journal{Journal of \LaTeX\ Templates}

\biboptions{authoryear}

\journal{Advances in Space Research}

\newcommand{\dif}{\mathrm{d}}

\begin{document}

\begin{frontmatter}

\title{Model of a fluxtube  with a twisted magnetic field in the stratified solar atmosphere}
\author[iia]{S.\ Sen\fnref{eid1}}
\author[iia]{A.\ Mangalam \fnref{eid2} \corref{cor}}
\address[iia]{Indian Institute of Astrophysics, Sarjapur Road, Koramangala, Bangalore 560034, India}
\fntext[eid1]{samrat@iiap.res.in}
\fntext[eid2]{mangalam@iiap.res.in}
\cortext[cor]{Corresponding author}

\begin{abstract}

We build a single vertical straight magnetic fluxtube  spanning the solar photosphere and the transition region which does not expand with height. We assume that the fluxtube containing twisted magnetic fields  is in magnetohydrostatic equilibrium within a realistic stratified atmosphere subject to solar gravity. Incorporating specific forms of current density and gas pressure in the Grad--Shafranov equation, we solve the magnetic flux function, and find it to be separable  with a Coulomb wave function in  radial direction while the vertical part of the solution decreases exponentially. We employ improved fluxtube boundary conditions  and  take a realistic ambient external pressure for the photosphere to transition region, to derive a family of solutions for reasonable values of the fluxtube radius and  magnetic field strength at the base of the axis that are the free parameters in our model. We find that our model estimates are consistent with the magnetic field strength and the radii of Magnetic bright points (MBPs) as estimated from observations. We also derive thermodynamic quantities inside the fluxtube.

\end{abstract}


\end{frontmatter}


\section{Introduction}

The study of small scale magnetic structures in the solar photosphere is  important  because they play a crucial role in the evolution of  active regions and  sunspots \citep{muller1987,2007ApJ...666L.137C}.  Magnetic bright points (MBPs) are likely to be the fluxtubes observed in the photosphere \citep{1995ApJ...454..531B,2007ApJ...666L.137C,2010ApJ...723L.164L}. 
The topological rearrangement of these  magnetic fluxtubes  due to the motion of the photospheric foot points or magnetic reconnections,  contribute to the coronal heating  \citep{1994A&A...283..232M,1986ApJ...311.1001V}. A three dimensional (3D) single fluxtube model with untwisted magnetic field has been studied by solving linear elliptic partial differential equation by numerical iterative process \citep{1986A&A...170..126S}. \cite{Schluter_Temesvary_1958} and \cite{1984SoPh...94..207O} studied a 3D fluxtube for sunspots using a self--similar model. The magnetic and thermodynamic structure inside fluxtube with untwisted magnetic field which spans from photosphere to the lower part of the solar corona is studied by \cite{2013MNRAS.435..689G}. Both 2D and 3D numerical models of fluxtubes  with the energy propagation through the torsional Alfven waves have been studied by \cite{2015SoPh..290.1909M,2015A&A...577A.126M}, where an empirical form of magnetic flux function was assumed. \cite{2009A&A...508..951V} assumed an empirical form of gas pressure to investigate the wave propagation and energy transport through a fluxtube. Several interesting results of wave behavior in the solar photosphere and chromosphere have been presented by several authors \citep{2003ApJ...599..626B,2009SoPh..258..219F,2010A&A...515A.107S}.\\
In this work,  we construct a 3D single cylindrical vertical straight magnetic fluxtube semi-analytically with a twisted magnetic field by obtaining a new solution of poloidal  flux function by solving Grad--Shafranov equation (GSE; \cite{1958conf...21...190,1958jetp...710...33}). We assume a specific form of gas pressure and poloidal current, which has been used to study the equilibrium solution of terrestrial plasma  \citep{doi:10.1063/1.1756167}. An equilibrium solution near the magnetic axis of the plasma torus has been reported previously, using  a plasma pressure  and poloidal current profile that varies linearly with the poloidal  flux function \citep{1968JETP...26..2S}. We obtain an analytic solution by assuming a form that is quadratic  in the poloidal  flux function, and derive the magnetic field structure and thermodynamic quantities inside the fluxtube using the solution that represents an ideal MHS equilibrium.  In the future, we will look to explore fully the profile functions that will improve the solution set.\\
The overview of the paper is as follows. In section 2, the GSE has been derived assuming a specific form of the profile function of gas pressure and poloidal current  and the solution of the equation is presented. In section 3, we discuss the boundary condition that is  physically acceptable, and can be used for realistic modelling of a fluxtube. In section 4, the mode wise variation of the profile functions are presented and in section 5, we compare the model with the observations. Finally, we conclude with a comparison  with other existing models. 

\section{Solution of Grad-Shafranov equation}

We assume an axisymmetric cylindrical geometry, with gas pressure $p$ and take the poloidal current $I_p$  constant along   a magnetic field line. We express $p(\Psi,z)$ and $I_p(\Psi)$ in terms of the poloidal  flux function $\Psi(r,z)$  and $z$ and consider a straight vertical axisymmetric fluxtube  that spans the altitude from photosphere ($z=0$) to the transition region ($z=2.15$ Mm) that is in  equilibrium  with the atmosphere outside with the uniform gravity  ${\bf g} (=-g \hat{z})$  acting vertically downward. The force balance equation in MHS equilibrium takes the form
\begin{equation}\label{mhs}
-\nabla p+\frac{1}{4\pi}(\nabla \times {\bf B})\times {\bf B}+\rho {\bf g}=0, 
\end{equation}
where $\rho$ denotes the mass density and ${\bf B}$ is the magnetic field associated with the poloidal  flux function $\displaystyle{\Psi(r,z)= \int_0^{r}B_z(r',z) r' {\rm d}r' }$ (scaled by the factor $\frac{1}{2\pi}$) in the following form
\begin{align}\label{mc}
B_r=-\frac{1}{r}\frac{\partial \Psi}{\partial z} ;\quad B_z=\frac{1}{r}\frac{\partial \Psi}{\partial r} ; \quad B_\phi=\frac{I_p}{r}.
\end{align}
This form of $B_r,B_\phi$ and $B_z$  ensures the solenoidal condition of magnetic field. Now splitting the MHS force balance equation (\ref{mhs}) into $r$ and $z$ directions, we find two different scalar partial differential equations
\begin{subequations}
\begin{align}\label{gsec}
-\frac{\partial p}{\partial r}+\frac{1}{4\pi}\bigg(B_r\frac{\partial B_r}{\partial r}+B_z\frac{\partial B_r}{\partial z}-\frac{1}{2}\frac{\partial B^2}{\partial r}\bigg)=0\\ \label{gsec2}
-\frac{\partial p}{\partial z}+\frac{1}{4\pi}\bigg[\frac{1}{r}\frac{\partial \Psi}{\partial z}\bigg(\frac{1}{r^2}\frac{\partial \Psi}{\partial r}-\frac{1}{r}\frac{\partial^2\Psi}{\partial r^2}\bigg)-\frac{1}{r^2}\frac{\partial \Psi}{\partial z}\frac{\partial^2\Psi}{\partial z^2}-\frac{1}{2r^2}\frac{\partial}{\partial z}(I_p^2)\bigg]-\rho g=0.
\end{align}
\end{subequations}
 If the gas pressure and poloidal current are functions of $\Psi$ alone i.e., $p_1(\Psi)$ and $I_p(\Psi)$ respectively, then from the (\ref{gsec}, \ref{mc}) it follows that
\begin{align}\label{rc}
\frac{\partial^2\Psi}{\partial r^2}-\frac{1}{r}\frac{\partial \Psi}{\partial r}+\frac{\partial^2\Psi}{\partial z^2}=-\frac{1}{2}\frac{\partial I_p^2(\Psi)}{\partial \Psi}-4\pi r^2\frac{\partial p_1(\Psi)}{\partial \Psi}.
\end{align}
Plugging in $p_1$ and $I_p^2$ in  (\ref{gsec2}), we find
\begin{align}\label{zc}
-\frac{\partial p_1(\Psi)}{\partial z}+\frac{1}{4\pi}\bigg[\frac{1}{r^2}\frac{\partial \Psi}{\partial z}\bigg(\frac{1}{r}\frac{\partial \Psi}{\partial r}-\frac{\partial^2\Psi}{\partial r^2}\bigg)-\frac{1}{r^2}\frac{\partial \Psi}{\partial z}\frac{\partial^2\Psi}{\partial z^2}-\frac{1}{2r^2}\frac{\partial I_p^2(\Psi)}{\partial z}\bigg]-\rho g=0.
\end{align}
By multiplying both sides of (\ref{zc}) by $\displaystyle{4\pi r^2 \frac{\partial z}{\partial \Psi}}$ and using (\ref{rc}), we obtain $\displaystyle{g \rho \frac{\partial z}{\partial \Psi}=0}$ which implies that  $\rho$ is zero, which means that the vertical hydrostatic pressure balance will not be maintained. Therefore, to balance the vertical hydrostatic pressure  inside the fluxtube, we introduce a new function, $p_2(z)$ such that, \begin{displaymath} p(r,z)=p_1(\Psi)+p_2(z). \end{displaymath}  We assume  $p_1(\Psi)$ and $I_p^2(\Psi)$ to be second order polynomials of $\Psi$
\begin{subequations}
\begin{align}\label{p}
p(r,z)= p_1(\Psi)+p_2(z) \\ \label{bphi1}
I_p^2(r,z)=\tilde{\alpha} \Psi^2+ \tilde{\beta} \Psi+I_0^2
\end{align}
\end{subequations}  
where \begin{displaymath} p_1(\Psi)=\tilde{a} \Psi^2+\tilde{b} \Psi, \end{displaymath} and the parameters $\tilde{a}, \tilde{b}, \tilde{\alpha}, \tilde{\beta}$ and $I_0^2$ are to be determined by appropriate boundary conditions. The function $p_2(z)$ is to be evaluated later.  The substitution of $p$ given by (\ref{p}) in (\ref{gsec}) gives  (\ref{rc}) and we obtain the following second order scalar partial linear inhomogeneous differential equation
\begin{align}\label{gse2}
\frac{\partial^2\Psi}{\partial r^2}-\frac{1}{r}\frac{\partial \Psi}{\partial r}+\frac{\partial^2\Psi}{\partial z^2}= -(ar^2+\alpha)\Psi-(br^2+\beta),
\end{align}
with the rescaled parameters, $\displaystyle{a=8\pi \tilde{a}; \alpha=\tilde{\alpha}; b= 4\pi \tilde{b}; \beta=\tilde{\beta}/2 }$. To solve  (\ref{gse2}), we split $\Psi$ in two parts:  a homogeneous part, $\Psi_h(r,z)$ and an inhomogeneous part $\Psi_p(r)$, i.e. $\Psi(r,z)=\Psi_h(r,z)+\Psi_p(r)$. Using this form in  (\ref{gse2}), we separate the homogeneous and the inhomogeneous parts to obtain the following expressions:
\begin{subequations}
\begin{align}\label{hom}
\frac{\partial^2\Psi_h}{\partial r^2}-\frac{1}{r}\frac{\partial \Psi_h}{\partial r}+\frac{\partial^2\Psi_h}{\partial z^2}= -(ar^2+\alpha)\Psi_h  \\ \label{inhom}
\frac{\partial^2\Psi_p}{\partial r^2}-\frac{1}{r}\frac{\partial \Psi_p}{\partial r}= -(ar^2+\alpha)\Psi_p-(br^2+\beta).
\end{align}
\end{subequations}
To solve the homogeneous part, we seek a solution of the form $\Psi_h(r,z)= S(r) Z(z)$. Then we separate out the $r$ and $z$ part in (\ref{hom}) as follows
\begin{align}\label{separate}
\frac{S''}{S}-\frac{1}{r}\frac{S'}{S}+a r^2+\alpha=-\frac{Z''}{Z}=-k^2,
\end{align}
where $k$ is an arbitrary real constant. Motivated by the fact that the poloidal  flux function $\Psi(r,z)$ decreases with  $z$,  we assume that the solution of the $z$-part of (\ref{separate}) takes the form
\begin{align}
Z(z)=C e^{-k z} ,
\end{align}
where $C$ is an arbitrary constant. To solve the $r$--part of (\ref{separate}), we substitute $x=\frac{\sqrt{a} r^2}{2}$ (where $a>0$) and insert it in  (\ref{separate}) to find
\begin{align}\label{coulomb_eq}
\frac{{\rm d}^2S}{{\rm d}x^2}+\bigg(1+\frac{2 \eta}{x}\bigg)S=0 ,
\end{align}
whose solutions are given by Coulomb wave functions $F_L(-\eta,x)$ and $G_L(-\eta,x)$ \citep{abramowitz+stegun}  (page $537-544$) with $L=0$ and $\displaystyle{\eta=\frac{\alpha+k^2}{4\sqrt{a}}}$. The solution of (\ref{coulomb_eq}) takes the following form
\begin{align}\label{S}
S(r)= C_1 F_0\big(-\eta,\frac{\sqrt{a}r^2}{2}\big)+C_2  G_0\big(-\eta,\frac{\sqrt{a}r^2}{2}\big).
\end{align}
Here  $F_0\big(-\eta,\frac{\sqrt{a}r^2}{2}\big)$ and $G_0\big(-\eta,\frac{\sqrt{a}r^2}{2}\big)$ are called the regular and irregular Coulomb wave functions respectively which are complex quantities  with real arguments \citep{1968ams...2B}, given by
\begin{align}\label{gwf}
F_0(-\eta,\frac{\sqrt{a}r^2}{2})=C_0(\eta) M_{-i \eta,1/2}(i \sqrt{a}r^2), \\
G_0(-\eta,\frac{\sqrt{a}r^2}{2})=i C_0(\eta) M_{-i \eta,1/2}(i \sqrt{a}r^2) + D_0(\eta) W_{-i \eta,1/2}(i \sqrt{a}r^2),
\end{align}
where $ M_{-i \eta,1/2}(i \sqrt{a}r^2)$ and $ W_{-i \eta,1/2}(i \sqrt{a}r^2)$ are called the Whittaker-M and Whittaker-W function  (see Figure \ref{Ar} below) and the constants $C_0(\eta)$ and $D_0(\eta)$ are defined by 
\begin{align}
C_0(\eta)=\frac{1}{2}|\Gamma(1-i\eta)| e^{-\frac{\pi}{2}(i-\eta)} 
\end{align}
and
\begin{align}
D_0(\eta)=\frac{\Gamma(1-i\eta)}{|\Gamma(1-i\eta)|} e^{-\frac{\pi \eta}{2}}.
\end{align}
The Whittaker function has been used by several authors in their models of the solar atmosphere albeit in different physical problems (eg. \cite{1979ApJ...231..260T} in the context of inviscid flows, and also in the context of MHD waves by \cite{2008ApJ...677..769H} and \cite{2010SoPh..263...63E}). Now $B_z(r,z)$ has to be a finite quantity that varies linearly with the term $\displaystyle{\frac{1}{r} \frac{\dif S}{\dif r}}$ but  $\displaystyle{\frac{1}{r} \frac{\dif}{\dif r}[G_0(-\eta,\frac{\sqrt{a} r^2}{2})]}$ blows up at $r=0$; therefore for $B_z$ to be finite  on the axis of the fluxtube $C_2$ in  (\ref{S}) must vanish. As a result  $S(r)$ takes the form              
\begin{align}\label{S2}
S(r)= C_1 F_0\big(-\eta,\frac{\sqrt{a}r^2}{2}\big) ,
\end{align}
and the homogeneous part of the solution is given by
\begin{align}\label{Ah}
\Psi_h(r,z)=C e^{-kz} F_0\big(-\eta,\frac{\sqrt{a}r^2}{2}\big).
\end{align}
A similar but a different solution, which is oscillatory in $z$ is used for laboratory plasma  for both a $D$-shaped plasma and toroidally diverted plasma \citep{doi:10.1063/1.1756167}. The general solution of  (\ref{gse2}) is given by the sum of the homogeneous part $\Psi_h(r,z)$ given above and an inhomogeneous part $\Psi_p(r)$  which will be presented in a paper in preparation. We have found that the presence of $\Psi_p(r)$ term in the poloidal flux function $\Psi(r,z)$, implies that $p$ and $I_p^2$ cannot be simultaneously positive for any combination of $b$ and $\beta$ in the physical parameter domain space for all $r$ and $z$. For avoiding these unphysical effects we present the case of $\Psi=\Psi_h$ and an exploration of the general solution $\Psi=\Psi_h+\Psi_p$ will be studied  later. Since $\Psi(r,z)$ and its complex conjugate function, $\Psi^*(r,z)$ are the valid solutions of  (\ref{gse2}), we construct  a solution of (\ref{gse2})  by redefining  $\displaystyle{\frac{\Psi(r,z)+\Psi^*(r,z)}{2} \rightarrow \Psi(r,z)} \equiv \varsigma(r) Z(z)$.


\section{Boundary conditions and the reduced form of $p$ and $I_p$}

The ideal magnetic fluxtube is embedded in a magnetic field free region with no current outside the fluxtube boundary. We make the following standard assumptions $B_r(r=R,z)=0$ and  $B_\phi(r=R,z)=0$ to ensure that there is no net current $I_p$  at the fluxtube boundary. The pressure at the photosphere ($z=0$) outside the fluxtube  is $p_0=1.228 \times 10^5$ dyne cm$^{-2}$ and at the transition region ($z_{tr}=2.15$ Mm) is $p_{tr}=0.1058$ dyne cm$^{-2}$  and is taken from Avrett-Loeser model \citep{2008ApJS..175..229A}. We summarize the boundary conditions below
\begin{subequations}
\begin{align}\label{bc1}
B_r(R,z)=0\\ \label{bc2}
B_\phi(R,z)=0 \\ \label{bc3}
p(R,0)=p_0 \\ \label{bc4}
p(R,z_{tr})=p_{tr} 
\end{align}
\end{subequations}
Assuming that pressure decreases exponentially from photosphere to transition region, we use the following expression for the external pressure
\begin{align}\label{pex}
p_{ex}(z)=p_0 \exp(-2kz),
\end{align}
where $\displaystyle{k=\frac{1}{2 \times 2.15} \ln\big(\frac{p_0}{p_{tr}}\big)}$ Mm$^{-1}=3.248$ Mm$^{-1}.$ Matching the pressure scale heights,  we see that $p_2(z)$ (\ref{p}) also decreases exponentially with $z$ as
\begin{align}\label{p2}
p_2(z)=p_{20}\exp(-2kz),
\end{align}
where $p_{20}$ will need to be calculated. Taking $\Psi(r,z)=\Psi_h(r,z),$ the reduced forms of $p$ and $I_p^2$ are given by
\begin{align} \label{gp}
p(\Psi,z)&=\frac{a}{8\pi} \Psi^2+p_2(z)  \quad (a>0)
\end{align}
and
\begin{align} \label{bphi2}
I_p^2(\Psi)&=\alpha \Psi^2 \quad (\alpha>0).
\end{align}    
Taking the radial component of the MHS force balance equation (\ref{mhs}) and adding the contribution of the radial force due to the presence of sheet current $j_\phi$ at the boundary we write the force balance equation
\begin{align}\label{fbe}
-\frac{\partial p}{\partial r}\bigg|_{r=R}+\frac{1}{4\pi}\bigg(B_r\frac{\partial B_r}{\partial r}+B_z\frac{\partial B_r}{\partial z}\bigg)\bigg|_{r=R} -\frac{\partial}{\partial r}\bigg(\frac{B^2}{8\pi}\bigg)\bigg|_{r=R}+j_\phi(r)B_z(r)\bigg|_{r=R}=0.
\end{align}
Now the sheet current $j_\phi$ can be expressed as a delta function $j_\phi(r)=j_{\phi s} \delta(r-R)$ which is non zero only at the boundary. Integrating  (\ref{fbe}) w.r.t. $r$ from $r=R-\epsilon$ to $r=R+\epsilon$ where $\epsilon$ is an infinitesimal positive quantity we obtain

\begin{figure}
\begin{center}
\includegraphics[scale=0.5]{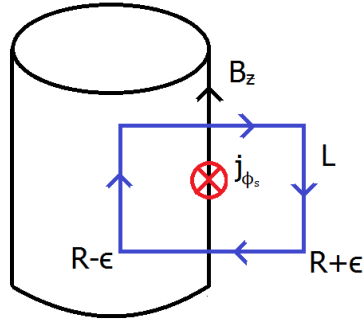}
\end{center}
\caption{Fluxtube geometry at the boundary showing the sheet current.}
\label{tube}
\end{figure}

\begin{align}
-\int_{R-\epsilon}^{R+\epsilon}\frac{\partial p}{\partial r}dr+\frac{1}{4\pi}\bigg(\int_{R-\epsilon}^{R+\epsilon}B_r\frac{\partial B_r}{\partial r}dr+\int_{R-\epsilon}^{R+\epsilon}B_z\frac{\partial B_r}{\partial z}dr\bigg)\\ \nonumber -\int_{R-\epsilon}^{R+\epsilon}\frac{\partial}{\partial r}\bigg(\frac{B^2}{8\pi}\bigg)dr+ \int_{R-\epsilon}^{R+\epsilon}j_\phi(r)B_z(r){\rm d}r=0,
\end{align}
which leads to the MHS force balance at the boundary to be given by
\begin{align}\label{fbe2}
p_{in}-p_{ex}+j_{\phi s}B_z(R)+\frac{1}{4\pi}\bigg[ B_r \frac{\partial B_r}{\partial r}+B_z\frac{\partial B_r}{\partial z} \bigg]_R+\frac{B_i^2-B_e^2}{8\pi}=0 ,
\end{align} 
where $[...]_R$ denotes the jump condition at the boundary and $B_i$ and $B_e$ are the magnetic fields inside and outside the fluxtube boundary. This is an improved  boundary pressure condition for a magnetic fluxtube, as previous studies have ignored the sheet current. Now to calculate $j_{\phi s}$, we assume an infinitesimal current loop at the boundary which has a vertical height of length $L$ and radial extent from $R-\epsilon$ to $R+\epsilon$.  Using Stokes line integral theorem along the closed loop (see Fig.\ref{tube}),  we see that 
\begin{align}
B_z L=4\pi L \int_{R-\epsilon}^{R+\epsilon}j_{\phi s}\delta(r-R){\rm d}r,
\end{align} 
which implies $\displaystyle{j_{\phi s}=\frac{B_z}{4 \pi}}$. Since $B_r(R)=B_\phi(R)=B_e(R)=0$ for any height $z$ and $p_{in}(R,z)=p_2(z)$, from  (\ref{fbe2}), the total pressure at the boundary inside the fluxtube is $\displaystyle{p_2(z)+\frac{3B_z^2(R)}{8\pi}}$ and matching the pressures gives
\begin{align}\label{p21}
p_2(z)=p_{ex}(z)-\frac{3B_z^2(R,z)}{8\pi}.
\end{align}
The mass density inside the fluxtube obtained from  (\ref{gsec2}) is 
\begin{align}\label{density}
\rho(z)=-\frac{1}{g}\frac{\dif p_2(z)}{\dif z},
\end{align}
and the density inside the fluxtube varies only with $z$ and at the transition region ($z_{tr}=2.15$ Mm), which should match with the external density which is typically $\rho_{tr}=2.77 \times 10^{-14}$ g cm$^{-3}$ \citep{2008ApJS..175..229A}. From  (\ref{p21}) and  (\ref{pex}) we see that 
\begin{align}
p_{20}=\frac{g}{2k} e^{2 k z_{tr}} \rho_{tr} .
\end{align}
In our model $g$ is assumed not to vary much from photosphere to the transition region and  its value is taken to be  $g=274$ m s$^{-2}$, the value at the solar surface and that determines $p_{20}=1.36\times 10^4$ dyne cm$^{-2}$. Using the forms of $B_z(r,z)$ from (\ref{mc}, \ref{Ah}) we obtain $B_z(0,0)=4 \sqrt{a}C \equiv B_{z0}$, and the relation between $C,B_{z0}$ and $a$ is derived in the following.\\ 
The real component of $\Psi$ is given by, \begin{displaymath} {\rm Re}(\Psi)=\frac{\Psi+\Psi*}{2}
\end{displaymath} and by using  (\ref{gwf}) and  (\ref{Ah}) we can express the flux function in the form of Whittaker--M functions as 
\begin{align}\label{a1}
\Psi(r,z)= C e^{-k z} \bigg[M_{-i \eta,1/2}(i \sqrt{a} r^2)+M_{i \eta,1/2}(-i \sqrt{a} r^2)\bigg].
\end{align} 
Whittaker--M function can be expressed in terms of hypergeometric function by the standard relation \citep{2015msa...26...49}
\begin{align}
M_{t,m}(z)=e^{-z/2} z^{m+1/2} F_1^1(1/2+m-t,1+2m,z).
\end{align}
Here $F_1^1$ represents the hypergeometric functions with the parameters $t,m$ and argument $z$. Therefore (\ref{a1}) takes the form 
\begin{align}
\Psi(r,z)=C e^{-kz} \sqrt{a}r^2 \bigg[e^{-i\sqrt{a}r^2/2} F_1^1(1+i\eta,2,i\sqrt{a}r^2)+e^{i\sqrt{a}r^2/2} F_1^1(1-i\eta,2,-i\sqrt{a}r^2)\bigg].
\end{align} 
It follows from (\ref{mc}) that $B_z$ takes the form 
\begin{align}\label{a2}
B_z(r,z)=\frac{\sqrt{a}C}{2}e^{-kz}e^{-i\sqrt{a}r^2/2}\bigg[(8+4i\sqrt{a}r^2)F_1^1\bigg(\frac{4\sqrt{a}+i(k^2+\alpha)}{4\sqrt{a}},2,i\sqrt{a}r^2\bigg)-\\ \nonumber r^2(4i\sqrt{a}+k^2+\alpha) F_1^1\bigg(\frac{4\sqrt{a}+i(k^2+\alpha)}{4\sqrt{a}},3,i\sqrt{a}r^2\bigg)\bigg].
\end{align}
So, $B_z(r,z)$ at $r=z=0$ is by definition $B_{z0}$ takes the form from (\ref{a2}) as,
\begin{align}
B_{z0}=4\sqrt{a}C F_1^1\bigg(\frac{4\sqrt{a}+i(k^2+\alpha)}{4\sqrt{a}},2,0\bigg).
\end{align}
But $F_1^1\bigg(\frac{4\sqrt{a}+i(k^2+\alpha)}{4\sqrt{a}},2,0\bigg)=1$  is an identity and therefore we finally have 
\begin{align}
C=\frac{B_{z0}}{4\sqrt{a}}.
\end{align} 

We define a physical observable $B_0$ which is the average magnetic field strength at the base  within the fluxtube as
\begin{equation}\label{B0}
B_0=\frac{1}{R}\int_0^R \sqrt{B_r^2(r,0)+B_z^2(r,0)+B_\phi^2(r,0)} {\rm d}r
\end{equation}
where $R$ is the radius of the fluxtube. Therefore we have two free parameters $R$ and $B_0$ that we can tune to fit our model with the observations. At $z=0$, from  (\ref{p21}), we get 
\begin{align}\label{p22}
\frac{3B_z^2(R,0)}{8\pi}=p_0-p_{20}.
\end{align}
Therefore, from (\ref{bc1}, \ref{bc2}, \ref{bc3}) we determine $a$, $\alpha$ and $C$ in terms of the free parameters $R$ and $B_0$ and hence the thermodynamic quantities within fluxtube. The temperature within the fluxtube is calculated by the ideal gas law according to the following form
\begin{align}\label{T}
T(r,z)=\frac{\bar{\mu} p(r,z)}{\rho(z) R_{g}},
\end{align}
where $R_{g}=8.314$ J mol$^{-1}$ K$^{-1}$ is the universal gas constant and 
\begin{displaymath}
\bar{\mu}=\frac{1}{z_{max}}\int_0^{z_{max}}\mu_{eff}(z) \dif z=1.116,
\end{displaymath}
is the average value of the mean effective molar mass from photosphere to transition region given by an empirical formula $\displaystyle{\mu_{eff}(z)=1.288 \big[ 1-0.535 \big( \frac{z}{2.152}\big)^3 \big]}$ \citep{2015AstL...41..211S} in the domain of $0<z<2.152$ Mm. A formulary of the different quantities are listed in Table \ref{function_tab}.

\begin{table}[h]
  \centering
  \resizebox{\textwidth}{!}{  
\begin{tabular}{|c | c |c |}\hline
Functions  & $r$--part & $z$--part \\ \hline
$\Psi(r,z)$ & $\varsigma(r)$ & $Z(z)$   \\  \hline
$B_r(r,z)$ & $\frac{3.248 \times 10^{-8}}{r} \varsigma(r)$ & $Z(z)$ \\  \hline
$B_\phi(r,z)$ & $\sqrt{\alpha} \frac{\varsigma(r)}{r}$ & $Z(z)$ \\  \hline
$B_z(r,z)$ & $\frac{\varsigma'(r)}{r}$ & $Z(z)$  \\  \hline
$p(r,z)$ & $\frac{a}{8\pi} \varsigma^2(r)+p_{20}$ & $Z^2(z)$  \\  \hline
$\rho(r,z)$ & $1$ & $(3.22 \times 10^{-8}) Z^2(z)$  \\  \hline
$T(r,z)$ & $0.0416\big(\frac{a}{8\pi}\varsigma^2(r)+p_{20}\big)$ & 1  \\  \hline
\end{tabular}
}
\caption{A formulary of the derived functions obtained from the solution of GSE. Here, $\varsigma(r)=C \big[F_0\big(-\eta,\frac{\sqrt{a}r^2}{2}\big)+F^*_0\big(-\eta,\frac{\sqrt{a}r^2}{2}\big)\big]$ and $Z(z)= e^{-kz}$. The value of the constants are $\bar{\mu}=1.116$, $g=2.74 \times 10^4$ cm s$^{-2}$, $k=3.248 \times 10^{-8}$ cm$^{-1}$, $p_{20}=1.36 \times 10^4$ dyne cm$^{-2}.$ All the quantities in the table are in cgs units.}
\label{function_tab}
\end{table}

\section{Mode analysis of different profile functions}

\begin{figure}
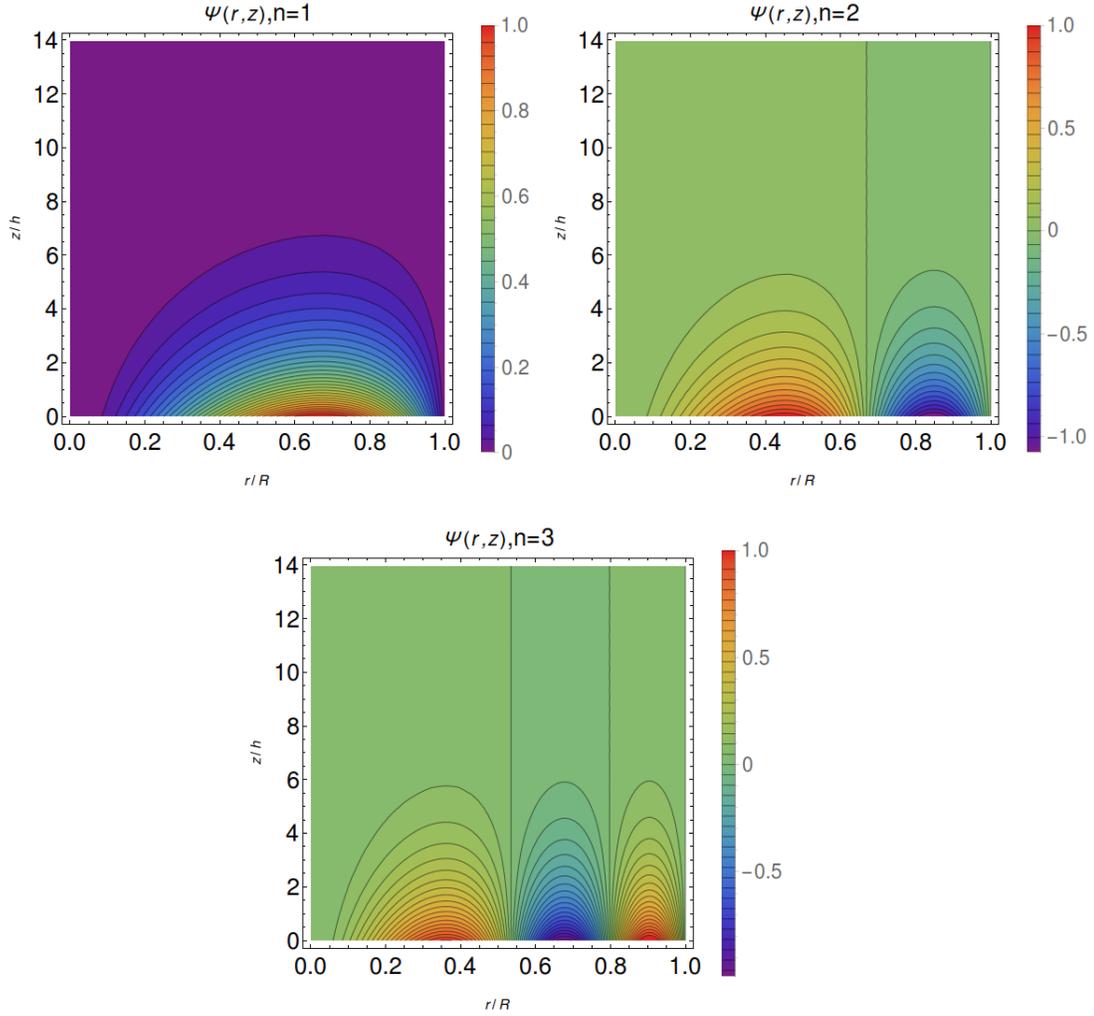

\includegraphics[scale=0.35]{2a.pdf}
\includegraphics[scale=0.35]{2b.pdf}
\begin{center}
\includegraphics[scale=0.35]{2c.pdf}
\caption{The vertical cross--sections of normalized poloidal  flux function for three different modes $n$ for $R=100$ km and $B_0=1$ kG are shown. The contours represent the magnetic lines of force in the $r-z$ plane. The amplitude of the flux function, normalized to the peak value, is represented by a colour bar. The horizontal axis is scaled to the radius of the fluxtube $R$ and the vertical axis is scaled with the pressure scale height, $h=162$ km.}
\label{A}
\end{center}
\end{figure}

\begin{figure}
\begin{center}
\includegraphics[scale=0.35]{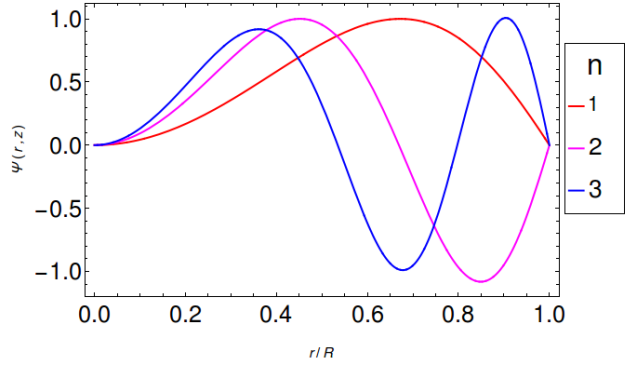}
\end{center}
\caption{The radial variations of the normalized poloidal flux function for three different modes for $R=100$ km and $B_0=1$ kG are shown.}
\label{Ar}
\end{figure}

\begin{figure}
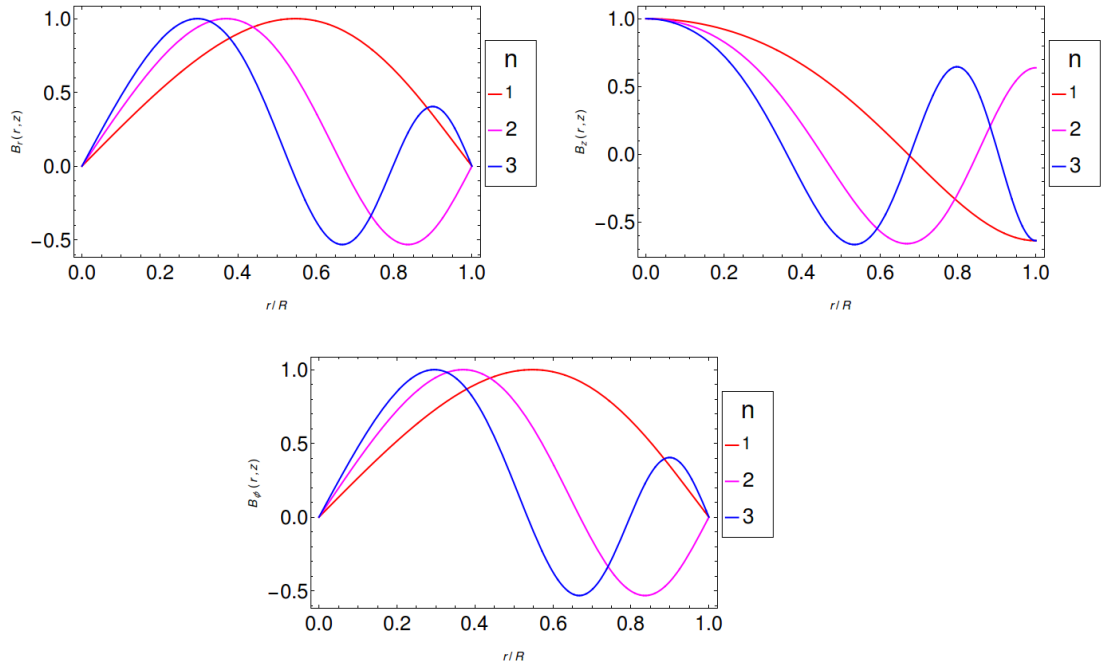

\includegraphics[scale=0.3]{4a.pdf}
\includegraphics[scale=0.3]{4b.pdf}
\begin{center}
\includegraphics[scale=0.3]{4c.pdf}
\caption{The radial variations of the normalized $B_r$, $B_\phi$ and $B_z$ fields for different modes for $R=100$ km and $B_0=1$ kG  are shown.}
\label{B1}
\end{center}
\end{figure}

\begin{figure}
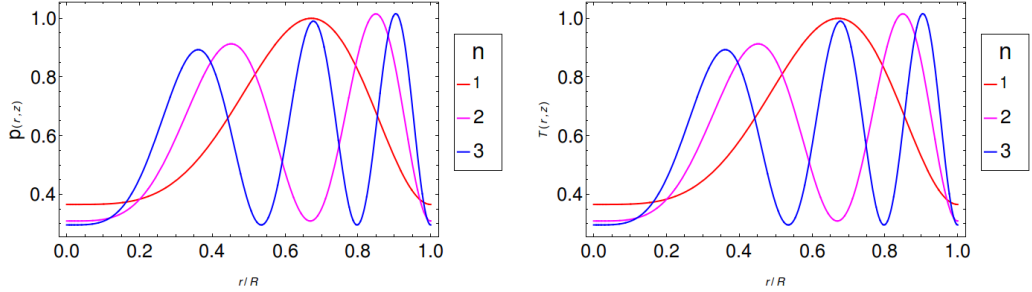

\includegraphics[scale=0.28]{5a.pdf}
\includegraphics[scale=0.28]{5b.pdf}
\caption{{\it Left:}  The radial variations of normalized gas pressure within the fluxtube are shown for three different modes $n$, {\it Right:}  The radial variations of normalized temperature profile inside the fluxtube are shown for three different modes $n$. Both the plots are for $R=100$ km and $B_0=1$ kG.}
\label{pt1}
\end{figure}

The quantities $a,\alpha$ and $C$ are functions of the free parameters $R$, $B_0$ and  the mode number $n$ whose values are given in Table \ref{parameter_mode} for a  sample set of the free parameters.     
\begin{table}[h]
  \centering
  \resizebox{\textwidth}{!}{  
\begin{tabular}{|c | c |c | c| c| c|}\hline
$R$ (km) & $B_0$ (kG) & mode no. & C ($10^{17}$ Mx)  & $\alpha$ ($10^{-14}$ cm$^{-2}$) & a ($10^{-28}$ cm$^{-4}$) \\ \hline
100 & 1 & 1 & 0.335061 & 7.50448 & 20.0417  \\  \hline
100 & 1 & 2 & 0.144828 & 15.6546 & 107.27 \\  \hline
100 & 1 & 3 & 0.0915463 & 24.0161 & 268.473 \\  \hline
\end{tabular}
}
\caption{Values of the  quantities $a,\alpha$ and $C$ for $R=100$ km and $B_0=1$ kG for three different modes.}
\label{parameter_mode}
\end{table}
The solutions to $\Psi, B, p$ and $T$ are shown for different mode numbers, in Figs. \ref{A}--\ref{pt1} respectively.

\begin{figure}
\begin{center}
\includegraphics[scale=0.5]{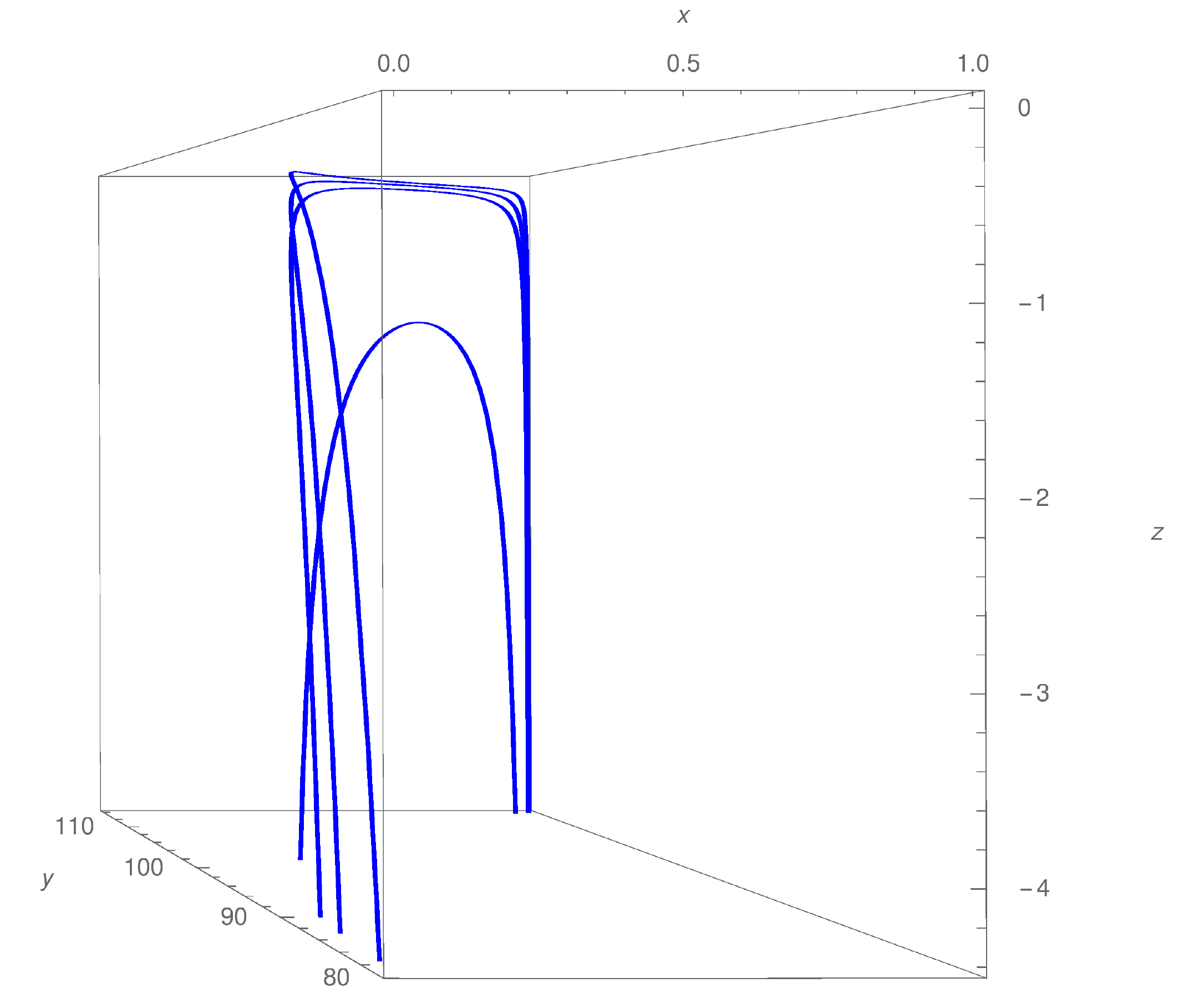}
\end{center}
\caption{The figure shows the $3$D topology of the magnetic field lines inside the fluxtube. The scales are in arbitrary units.}
\label{topology}
\end{figure}

All the profile functions are normalized to their peak values and the radial distance  to the total radius of the fluxtube $R$. As per the boundary conditions, the flux function, $\Psi$ vanishes both at the axis and at the boundary of the fluxtube, where the the total gas pressure is $p_2(z)$. The  solutions of higher modes have, the profile functions with higher frequency along the radial direction and realistically we may not have such reversible fields, as they are unstable. Therefore, we use the fundamental mode ($n=1$) for further analysis in the paper.  The $3$D topology of the magnetic field lines inside the fluxtube for the fundamental mode is shown in Fig. \ref{topology}.

\section{Comparing the model with observations}

Now we compare our model with the observations reported from high resolution and high cadence instruments. Small scale magnetic structures, {\it i.e.}, MBPs, are the best candidates for comparison because  such structures can be assumed to consist of fluxtubes. MBPs are seen in G-band filtergrams or are identified by making spectro-polarimetric measurements \citep{2009A&A...498..289U, 2013A&A...554A..65U, 2016SoPh..291.1089Y}.  The radial variation of the profile functions $p,\rho$ and $T$ and the magnetic components $B_r,B_\phi$ and $B_z$ are independent of $z$, but the amplitude  decreases exponentially with $z$ except for $T$. In the following, we validate the model by comparing the observed magnetic field strengths and radius of MBPs with those calculated in our model, and estimate the magnetic field strength and thermodynamic quantities at the transition region which may be verified by future observations.

\begin{figure}
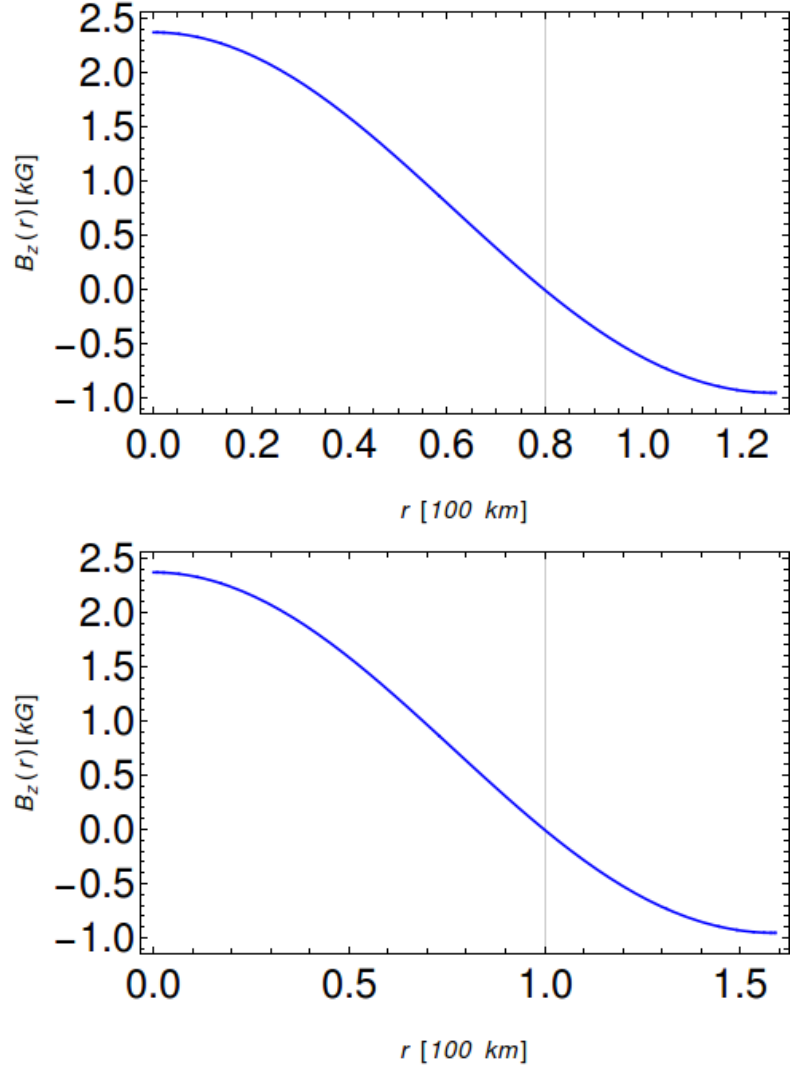

\begin{center}
\includegraphics[scale=0.6]{6a.pdf}
\includegraphics[scale=0.6]{6b.pdf}
\caption{The radial variations of the magnetic field strength at the photosphere ($z=0$) within the fluxtube for $R=127$ km (top) and $159$ km (bottom) are shown. The vertical grid line denotes the radius ($r_0$) beyond which $B_z$ becomes negative.  The values of $r_0$ are $80$ km and $100$ km for the top and bottom panels respectively. The horizontal axes are scaled in units of $100$ km and the vertical axes are scaled in units of kG for both top and bottom panels. The mean value of $B_z$ up to $r_0$ is $1.42$ kG in both panels.}
\label{Bz1}
\end{center}
\end{figure}

\begin{figure}
\begin{center}
\includegraphics[scale=0.75]{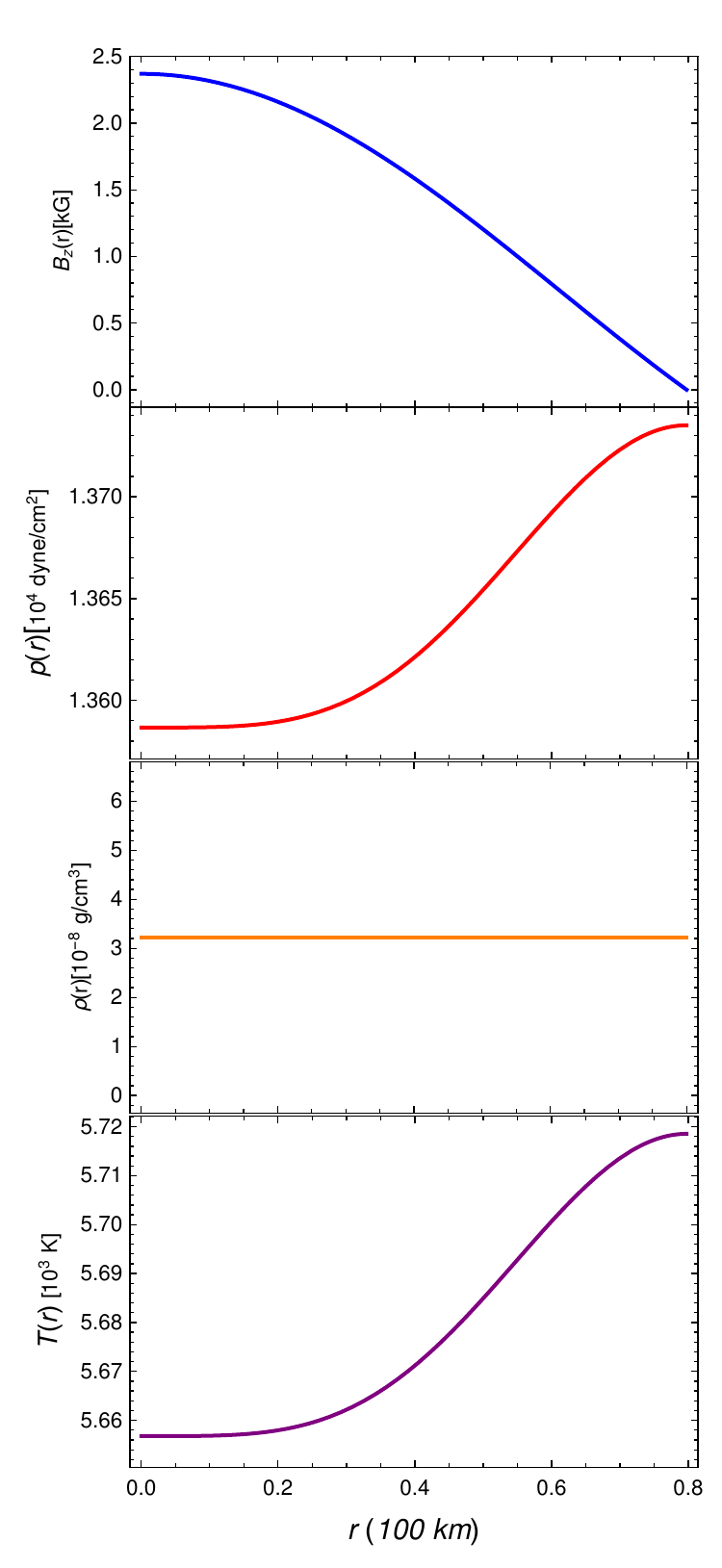}
\includegraphics[scale=0.75]{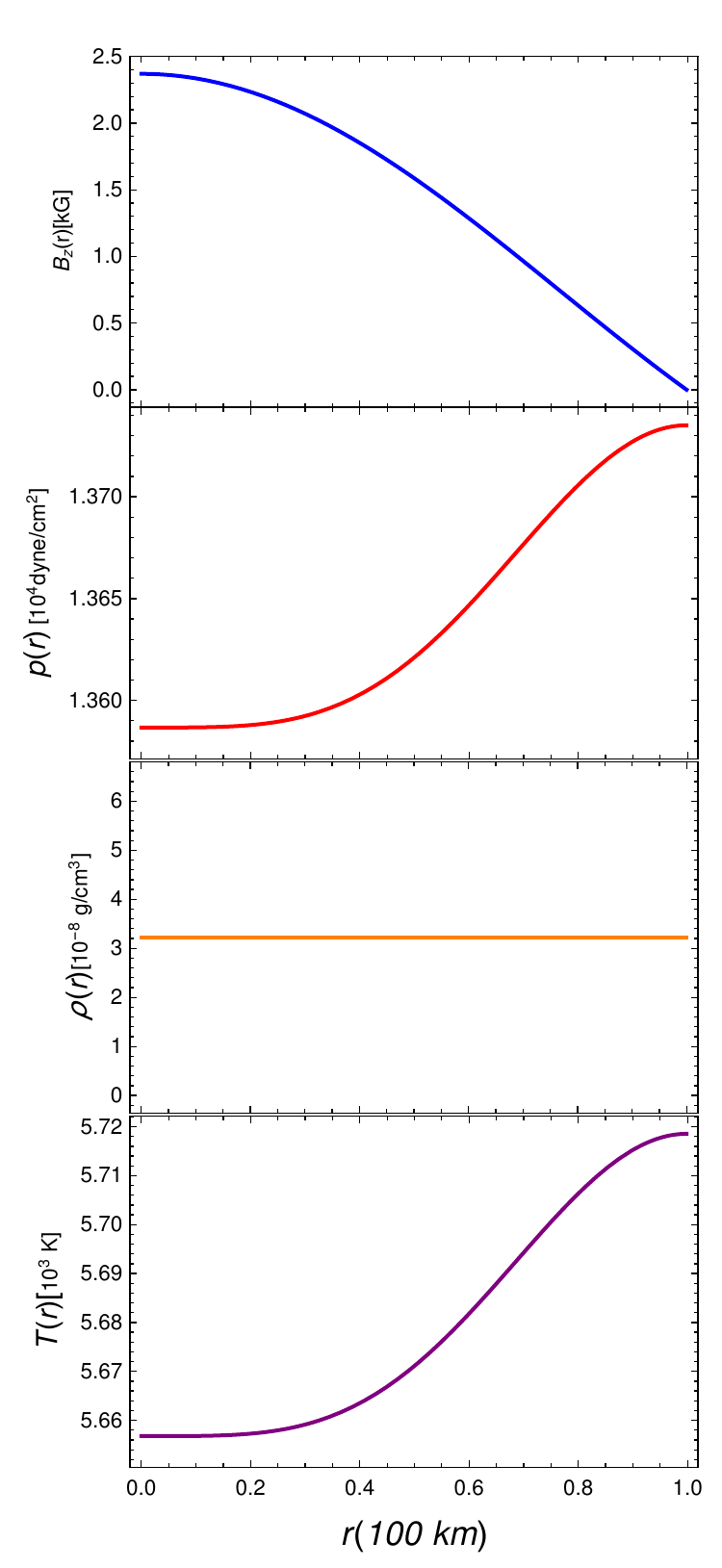}
\end{center}
\caption{From top to bottom: Predicted radial variation of vertical magnetic field strength, gas pressure, density and temperature from the axis to the boundary of the fluxtube  for two different sizes of MBPs ($80$ km radius in the left panel and $100$ km radius in the right panel). The horizontal axis is scaled in units of $100$ km and the vertical axes of $B_z, p, \rho, T$ are scaled in units of kG, $10^4$ dyne cm$^{-2}$, $10^{-8}$ g cm$^{-3}$ and $10^3$ K respectively.}
\label{BPT1}
\end{figure}

\begin{figure}
\begin{center}
\includegraphics[scale=0.9]{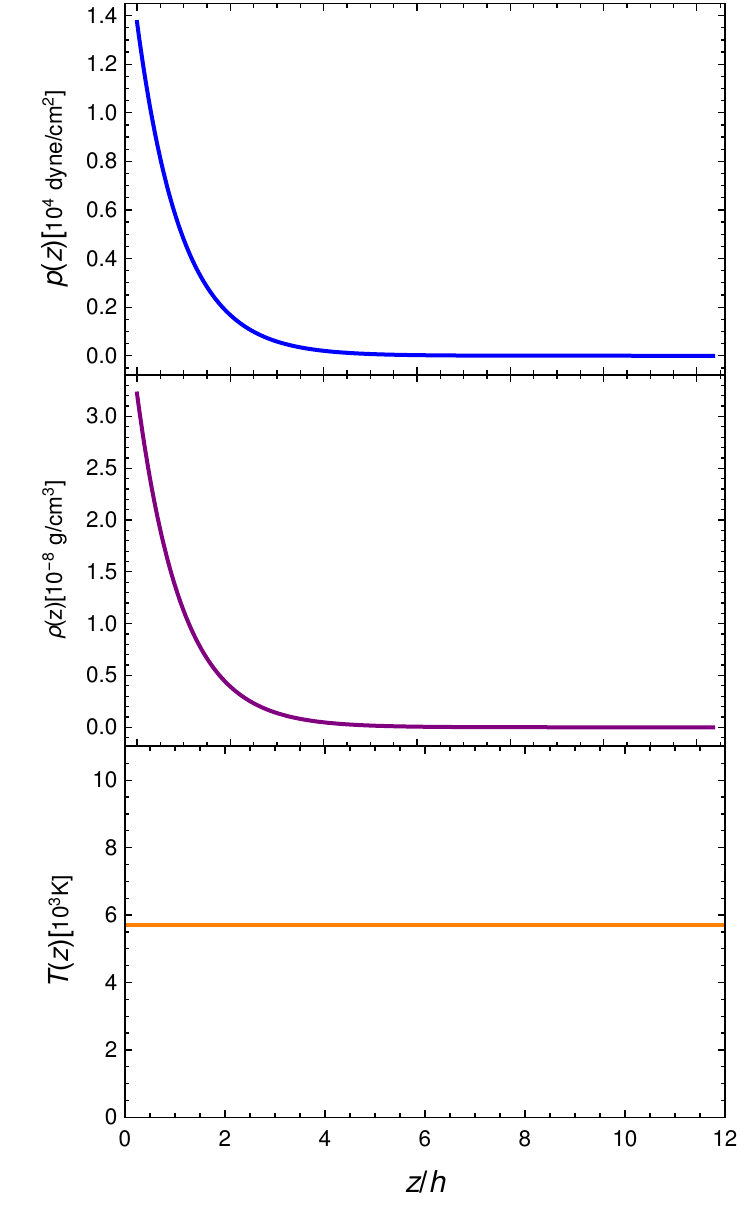}
\end{center}
\caption{The figure shows the variation of gas pressure, density and temperature at the BP boundary along $z$. The horizontal axis represents the height from the photosphere scaled with the pressure scale height $h=162$ km, and the vertical axes represents pressure, density and temperature from top to bottom respectively.} 
\label{thermo_fit} 
\end{figure}

The MBPs number distribution, magnetic field strength and  size distribution has been reported  by \cite{2009A&A...498..289U,2013A&A...554A..65U} at photosphere. The size distribution of MBPs peaks around $200$ km and $160$ km for low and high spatial sampling rates, respectively \citep{2009A&A...498..289U}. The magnetic field strength distribution is bimodal with two peaks at $\sim1400$ G and $\sim200$ G \citep{2013A&A...554A..65U}. Since MBPs are observed as the region of unipolar flux concentrations, we construct a cylindrical boundary inside the simulation domain where the vertical magnetic field $B_z$ is positive. We call this cut--off radius as $r_0$, where the line of sight magnetic field $B_z$ vanishes. The value of $B_z$ after this grid line  becomes negative. In Fig. \ref{Bz1}, the vertical grid line  denotes the boundary radius $r_0$. We study two different cases for $r_0=80$ and $100$ km which corresponds to the peak values for the MBP size distribution, for which $R$ is found to be $127$ and $159$ km respectively. 
For both cases, we calculate $B_0$ and the mean value of $B_z$, $\bar{B_z}$ in the radial direction up to $r_0$ and  find that for realistic values of the thermodynamic quantities inside the fluxtube, the upper limit of the vertical magnetic field strength $B_{z0}$ is $2.37$ kG. Beyond this value of $B_{z0}$, the viable solutions will shift to the higher modes. The temperature inside the fluxtube increases as the value of $B_{z0}$ decreases and temperature inside the fluxtube becomes greater than the typical photospheric temperature when $B_{z0}< 2.31$ kG. Thus it can be considered as the lower cut off limit of the magnetic field strength. The $\bar{B_z}$ value is only sensitive to $B_{z0}$ but not on $r_0$, and the thermodynamic quantities inside the fluxtube  remain the same for both $r_0=80$ and $100$ km. We found $\bar{B_z}=1.42$ and $1.4$ kG for $B_{z0}=2.37$ and $2.31$ kG respectively.  The radial and vertical variations of the vertical magnetic field strength, gas pressure, density and temperature inside the fluxtube are shown  in Figs. \ref{BPT1} and \ref{thermo_fit}. We see that the vertical magnetic field strength decreases from $2.37$ kG (on the axis at $z=0$) to zero at the MBP boundary ($r_0$, see Fig. \ref{Bz1}). The variation of gas pressure and temperature from axis to the MBP boundary is very small; at the photosphere the gas pressure changes $1.358\times 10^4$ (on the axis at $z=0$) to $1.373 \times 10^4$ dyne cm$^{-2}$ (at MBP boundary) and it decreases with $z$ to $3.12 \times 10^{-2}$ dyne cm$^{-2}$ (at MBP boundary) at the transition region ($z=2$ Mm).  The temperature changes from $5656$ K (on the axis) to $5718$ K (at MBP boundary)  which is small compared to the outside photosphere temperature ($6583$ K) \citep{2008ApJS..175..229A}. The average temperature inside fluxtube has been calculated by integrating the temperature from axis to the MBP boundary and is found to be $5679$ K. The density distribution is constant along the radius of the fluxtube at a given height which decreases with height from $3.22 \times 10^{-8}$ g cm$^{-3}$ at the photosphere to $7.33 \times 10^{-14}$  g cm$^{-3}$ at the transition region. The values of the quantities estimated from our modelled are summarized in Table \ref{thermo_tab}.

\begin{table}[h]
  \centering
  \resizebox{\textwidth}{!}{  
\begin{tabular}{|c | c |c | c| c| c|}\hline
$r$  & $z$ (Mm) & $B_z$ (G) & $p$ (dyne cm$^{-2}$)  & $\rho$ (g cm$^{-3}$) & $T$ (K) \\ \hline
0 & 0 & 2370 & $1.358 \times 10^4$ & $3.22 \times 10^{-8}$ & 5656  \\  \hline
0 & 2 & 2.19 & $3.09 \times 10^{-2}$ & $7.33 \times 10^{-14}$ & 5656 \\  \hline
$r_0$ & 0 & 0 & $1.373 \times 10^4$ & $3.22 \times 10^{-8}$ & 5718 \\  \hline
$r_0$ & 2 & 0 &$3.12 \times 10^{-2}$ & $7.33 \times 10^{-14}$ & 5718 \\ \hline
\end{tabular}
}
\caption{Table of the results obtained from our model for $r_0=80$ and $100$ km}
\label{thermo_tab}
\end{table}

\section{Conclusions and discussion}

In this paper, we constructed a single fluxtube with twisted magnetic field by solving GSE analytically. 
We summarize our results below:

\begin{enumerate}
\item We have an improved boundary condition by incorporating the sheet current  as compared to the previous studies e.g. \cite{2016SoPh..291.1647S}.

\item Our model depends on the form of the external pressure distribution  which is assumed as an exponentially decreasing function with $z$. Future observations leading to the more accurate form of pressure distribution from photosphere to transition region can be used to improve our model.  The plasma $\beta$ parameter inside the fluxtube remains constant with $z$ but it varies along $r$; $\beta <1$ is obeyed from the chromosphere to the transition region but not in the photosphere and lower atmosphere. Therefore the magnetic effects will dominate the gas dynamics throughout the simulation domain.

\item  In our model, the temperature varies along the radial direction, but it is constant along the vertical direction $z$. In other models e.g. \cite{2013MNRAS.435..689G} the  temperature rises with height from photosphere to the transition region and in \cite{2011SoPh..273...15V} the temperature decreases from $6300$ K at surface to $4000$ K (at $z=600$ km) and then it remains the same up to $1200$ km.

\item  The effects of shock wave dissipation and magnetic reconnection starts to dominate in the corona which causes the coronal heating. We have not considered these mechanisms in our model and therefore, we have not model the region in the corona or higher and have restricted our simulation domain to end at the transition region.


\item  Recently \cite{2014A&A...565A..84H, 2013MmSAI..84..369U, 2016arXiv161207887R}, have simulated bright points using MuRAM and Copenhagen-Stagger code.  We find that the magnitude of magnetic field strengths, pressure and densities reported in these studies are in fair agreement with our predictions but the temperature distribution along $z$ is in variance with  the results of the numerical simulations.

\end{enumerate}

The fluxtube model gives values of magnetic field and thermodynamic quantities consistent with observations and compares well with simulations. We plan to present a more detailed study of the model in a paper in preparation.

We thank Prof. P. Venkatakrishnan for useful discussions and the anonymous reviewers for insightful comments and helpful suggestions. We also thank the support staff of the IIA HPC facility and Sandra Rajiva for proofreading the manuscript.

\bibliography{fluxtube}

\end{document}